\documentclass[aps,prb,twocolumn,groupedaddress,showpacs]{revtex4}

\usepackage{graphicx}
\usepackage{dcolumn}
\usepackage{bm}
\usepackage{amsmath,amssymb}
\def\Z{{\mathbb{Z}}}

\begin{document}


\title{Quantum phase transitions of the asymmetric three-leg spin tube}
\author{T\^oru Sakai,$^{1,2}$ Masahiro Sato,$^{3}$  
Kouichi Okunishi,$^{4}$ Yuichi Otsuka,$^{2}$ 
Kiyomi Okamoto,$^{5}$ and Chigak Itoi$^{6}$ }
\affiliation{$^1$ Japan Atomic Energy Agency, SPring-8, Sayo, Hyogo 679-5148, Japan.\\
$^2$ Department of Material Science, University of Hyogo, Kamigori, 
   Hyogo 678-1297, Japan.\\
$^3$Condensed Matter Theory Laboratory, RIKEN, Wako, Saitama 351-0198, 
    Japan.\\
$^4$ Department of Physics, Niigata University, Niigata 950-2181, Japan.\\
$^5$ Department of Physics, Tokyo Institute of Technology, Meguro-ku, 
   Tokyo 152-8551, Japan.\\
$^6$Department of Physics, Nihon University, Kanda-Surugadai, Chiyoda-ku, 
  Tokyo 101-8308, Japan
}

\date{\today}

\begin{abstract}

We investigate
quantum phase transitions of the $S=\frac{1}{2}$ three-leg 
antiferromagnetic spin tube with asymmetric 
inter-chain (rung) exchange interactions. 
On the basis of the electron tube system, 
we propose a useful effective theory 
to give the global phase diagram of the asymmetric spin tube.
In addition, 
using other effective theories we raise the reliability of 
the phase diagram.
The density-matrix renormalization-group 
and the numerical diagonalization analyses show that
the finite spin gap appears in a narrow region around the rung-symmetric line,
in contrast to a recent paper by Nishimoto and Arikawa 
[Phys. Rev. B {\bf 78}, 054421 (2008)]. The numerical calculations indicate that 
this global phase diagram obtained by use of the effective theories 
is qualitatively correct. In the gapless phase on the phase diagram,
the numerical data are fitted by a finite-size scaling in 
the $c=1$ conformal field theory. 
We argue that all the phase transitions between the gapful and gapless 
phases belong to the Berezinskii-Kosterlitz-Thouless universality class. 
\end{abstract}

\pacs{75.10.Jm, 75.10.Pq, 75.30.Kz, 75.40.Cx}

\maketitle

\section{Introduction}

The spin tube,~\cite{nojiri,millet,garlea,zheludev,schulz,kawano,cabra,mila,fouet,
okunishi,sakai,arikawa,sato-sakai,sato07,sato05,sato-oshikawa} 
i.e., the spin ladder with the periodic boundary condition
along the inter-chain (rung) direction, is one of interesting magnetic
systems which is expected to exhibit some exotic phenomena due to 
special topology such as the carbon nanotube and the magnetic frustration. 
Actually, several theoretical 
works~\cite{kawano,mila,fouet,sakai,arikawa} 
have shown that the $S=\frac{1}{2}$ three-leg antiferromagnetic 
spin tube has a spin gap, in contrast to the corresponding three-leg
spin ladder with the open boundary condition along the rung. 
Moreover, it has been predicted in the field theoretical
method~\cite{sato-sakai,sato07} that both
the gapless and gapful vector-chiral long-range orders emerge in certain
parameter regions of the three-leg tube in a magnetic field. 
Recently some candidates for the spin nanotubes have been 
synthesized; 
a three-leg tube [(CuCl$_2$tachH)$_3$Cl]Cl$_2$
(Ref.~\onlinecite{nojiri}), 
a nine-leg tube ${\rm Na_2 V_3 O_7}$ (Ref.~\onlinecite{millet}), and 
a four-leg tube $\rm Cu_2Cl_4\cdot D_8C_4SO_2$ (Refs.~\onlinecite{garlea}
and \onlinecite{zheludev}). 
It is therefore expected that novel, intriguing phenomena will be 
detected in these materials.

Let us here focus on the $S=\frac{1}{2}$ three-leg
antiferromagnetic spin tube that is the simplest tube with geometrical
frustration. 
Since the unit cell consists of three spins with $S=\frac{1}{2}$ in this
tube, the Lieb-Schultz-Mattis theorem~\cite{lieb} suggests that 
the spin gap must be accompanied with at least doubly-degenerate ground states. 
In fact, previous numerical analyses~\cite{kawano,sakai,arikawa}
have confirmed such doubly-degenerate $S=0$ ground states 
due to the spontaneous breaking of the translational symmetry along the
leg direction. 
The ground states have a valence-bond type
(superposition of spin-singlet pairs) order.~\cite{kawano,arikawa} 
Here if one of the three rung coupling constants is changed,
the following two models are reproduced as limiting cases:
the three-leg spin ladder 
and the decoupled system of a single chain and a two-leg ladder.  
These two systems are believed to possess a gapless excitation. 
Therefore, the $S=\frac{1}{2}$ three-leg system, 
where one of the three rung couplings is varied, 
would undergo a quantum phase transition 
from the gapless state to the gapful symmetry-broken one. 
A recent numerical work~\cite{sakai} has suggested that
the gapful phase is extended to a finite (although narrow) region
when the rung-coupling asymmetry is introduced.
Unfortunately, however, 
the feature of the transition was not so clarified 
because the system size used in Ref.~\onlinecite{sakai} was too small. 
On the other hand, a recent density-matrix renormalization-group 
(DMRG) approach,~\cite{arikawa} 
assuming a special power-law form of the finite-size correction, 
has concluded that the transition is of the first-order and the system is 
always gapless except for the symmetric three-leg spin tube. 
Such a discontinuous transition, however, has not been 
reported so far in any realistic systems. 
In addition, from the viewpoint of the effective theory,
the occurrence of such a transition must 
require a highly fine tuning of parameters. 
Thus the phase diagram and the critical properties of the quantum phase
transitions in the three-leg spin systems are still controversial.

Motivated by the above situation, in this paper, 
we study the wide ground-state phase diagram and 
the universality classes of the quantum phase transitions 
in $S=\frac{1}{2}$ three-leg antiferromagnetic spin tube with the
rung-coupling asymmetry. We first propose a simple effective theory 
to explain the quantum phase transitions between gapful and gapless phases
on the basis of the Hubbard model on the tube lattice and the
non-Abelian bosonization.~\cite{affleck,tsvelik,gogolin} 
This effective theory enables us to draw a global phase diagram 
by counting the number of Fermi points. 
We find one gapful phase and three gapless phases 
in the phase diagram, and predict that the gapless phases are all 
described by a level-1 SU(2) Wess-Zumino-Witten (WZW) field 
theory,~\cite{affleck,tsvelik,gogolin} which is a $c=1$ conformal field
theory (CFT). Besides this effective theory, using other analytical 
strategies, we consider the strong-rung-coupling regime and 
the weak-rung-coupling regime with a strong rung distortion (i.e.,
asymmetry) in the phase diagram. 
In the former regime, two of three 
gapless phases are predicted by using 
a known effective theory.~\cite{arikawa}
In the latter regime, we prove that 
a finite gapless phase definitely exists.

We subsequently perform the numerical diagonalization 
and the DMRG calculation 
combined with some finite-size scaling analyses 
on the basis of the above effective theories. 
Applying the CFT approach~\cite{cft1,cft2,cft3} to our numerical data, 
we argue that the quantum phase transitions belong to the 
Berezinskii-Kosterlitz-Thouless (BKT) universality class.~\cite{bere,kt} 
A numerically quantitative phase diagram is presented, and 
is consistent with that of the effective theories.

This paper is organized as follows. 
In Sec.~\ref{sec:model}, we define the
Hamiltonian of an asymmetric three-leg spin tube model. 
In Sec.~\ref{effective}, we draw a qualitative but global
ground-state phase diagram by means of an effective theory 
based on the half-filled Hubbard model on the tube lattice. 
Employing another approach based on the non-Abelian bosonization, 
we precisely show the existence of
a gapless phase in the weak-rung-coupling regime with a strong
asymmetry. Moreover, we discuss the gapful phase in the
strong-rung-coupling regime. Section~\ref{numerical} is devoted to 
the numerical analyses for the spin tube. 
We plot a scaled gap calculated by the DMRG method to
confirm the phase diagram obtained by the effective theories.
We further analyze the numerical data obtained in the 
exact diagonalization on the basis of 
the finite-size scaling in the $c=1$ CFT.
We provide the summary and short discussions in Sec.~\ref{summary}.

\section{Model}
\label{sec:model}
\begin{figure}[h]
\begin{center}\leavevmode
\includegraphics[width=0.9\linewidth,angle=0]{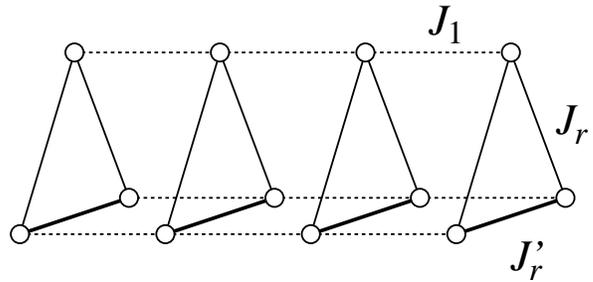}
\caption{Structure of the three-leg asymmetric spin tube~(\ref{ham}).}
\label{3tube}
\end{center}\end{figure}

We consider the $S=\frac{1}{2}$ asymmetric three-leg spin tube,
shown in Fig.~\ref{3tube}, 
described by the Hamiltonian 
\begin{eqnarray}
\label{ham}
H= &J_1& \sum _{i=1}^3 \sum_{j=1}^L \vec{S}_{i,j}\cdot \vec{S}_{i,j+1} \\
\nonumber
   &+&J_{\rm r} \sum _{i=1}^2 \sum_{j=1}^L \vec{S}_{i,j}\cdot \vec{S}_{i+1,j} 
   +J'_{\rm r} \sum_{j=1}^L \vec{S}_{3,j}\cdot \vec{S}_{1,j},
\end{eqnarray}
where $\vec{S}_{i,j}$ is the spin-$\frac{1}{2}$ operator and $L$ is the length of 
the tube in the leg direction. The exchange coupling constant $J_1$ 
is for the neighboring spin pairs along the legs, while $J_{\rm r}$ and
$J'_r$ are the rung coupling constants. 
All the exchange interactions are supposed to be antiferromagnetic (namely, positive). 
The ratio $\alpha=J'_r/J_{\rm r}$ stands for the degree of the asymmetry of
the rung couplings. 
We will vary $\alpha$ and $J_1$ to investigate the quantum phase transitions. 
Throughout this paper, we fix $J_{\rm r}$ to one. 

The present model includes three typical models as limiting cases;
(a) $\alpha=0$: the three-leg spin ladder, (b) $\alpha=1$: the symmetric spin tube, 
and (c) $\alpha \rightarrow \infty$: the single chain plus rung dimers. 
Since the system is gapless in the cases (a) and (c), while gapful in
the case (b), at least two quantum phase transitions should occur 
with increasing $\alpha$ from 0 to infinity. 
As we already mentioned, 
the one-site translational symmetry along the leg ($\vec S_{i,j}\to \vec S_{i,j+1}$)
is spontaneously broken in the symmetric spin tube
at least in the strong-rung-coupling regime.~\cite{kawano,arikawa}

\section{Effective theories \label{effective}}
In this section, we study the spin tube~(\ref{ham}) 
by constructing its low-energy effective theories. 
In Sec.~\ref{global}, we draw a phase diagram 
in the whole coupling-constant space $(\alpha, J_1)$.
To this end, we develop a simple effective theory for the spin tube 
from the corresponding Hubbard model on the tube lattice,
where the SU(2) spin-rotational symmetry is preserved automatically. 
This effective theory allows us to find three gapless phases and 
one extended gapful phase. 
Next, utilizing other theoretical schemes, 
we carefully investigate two special regimes 
$J_{\rm r} \ll J'_{\rm r}\ll J_1 $ and $J_{\rm r} \gg J_1 $, 
respectively, in Secs.~\ref{gapless_phase} and \ref{energy_gap}.

\subsection{Global phase diagram derived 
from the Hubbard model on the tube lattice}
\label{global}
Here, we provide a systematic 
method to draw global phase diagrams of one-dimensional 
antiferromagnetic quantum spin systems.
It is well-known that any $S=\frac{1}{2}$ Heisenberg model 
is obtained from the corresponding half-filled Hubbard model 
in the limit of strong on-site Coulomb interactions.
In one dimension, 
the spin configurations of the low-energy states
in the Heisenberg model qualitatively agree with those in 
the half-filled Hubbard model even with a weak Coulomb interaction. 
Furthermore, the phases in the weak-Coulomb regime often 
smoothly connect to those in the strong-Coulomb regime in
one-dimensional electron systems.  
Relying on these arguments, we construct the low-energy effective
theory for the spin tube~(\ref{ham}) from the corresponding Hubbard model. 
To discuss the wider parameter space, first
we diagonalize the kinetic parts of the Hubbard
Hamiltonian including both the leg and rung hopping
terms.~\cite{balents,Arrigoni,lin} 
Then, we take account of the on-site Coulomb interaction as 
the perturbation, with the help of the non-Abelian bosonization and CFT. 
As one will see later, the number of Fermi points is essential 
to determine whether or not the Coulomb interaction open a spin gap.

Since we consider the electron tube instead of the original spin tube, 
our results in this subsection should be regarded as a qualitative
argument. 
However, the approach from the electron tube can be applicable to the
wide parameter space $(\alpha,J_1)$, in contrast to the other
conventional methods. 
For example, the nonlinear sigma model is not usually reliable for 
frustrated magnets including the present spin tube. It is dangerous to
apply the Abelian bosonization to any SU(2)-symmetric magnets.~\cite{sato08} 
Moreover, a standard non-Abelian bosonization method, taking into
account rung couplings perturbatively (we will use it in
Sec.~\ref{gapless_phase}), 
is of course valid only in the weak-rung-coupling regime.  

Now, let us begin with the definition of the Hubbard model 
on the three-leg tube lattice. The Hamiltonian
\begin{equation}
\label{Hubbard_tube}
H=H_{\rm hop}+H_{\rm int},
\end{equation}
consists of the hopping part 
\begin{eqnarray}
H_{\rm hop} = \sum_{n=1} ^L \sum_{i=1}^3 \sum_{\sigma= \uparrow, \downarrow}
\hspace{-0.5cm} &&(t c_{n+1,i,\sigma}^\dag c_{n,i,\sigma}+ s_{i+1,i}  
c_{n,i+1,\sigma}^\dag c_{n,i,\sigma} \nonumber \\
&&+  {\rm h.c.}),
\end{eqnarray}
and the on-site interaction part
\begin{equation}
\label{coulomb}
H_{\rm int}= U \sum_{n=1} ^L \sum_{i=1}^3 n_{n,i,\uparrow} n_{n,i, \downarrow},
\end{equation}
where $n_{n,i,\sigma}=c_{n,i,\sigma}^\dag c_{n,i,\sigma}$ and $U > 0$ is
the repulsive coupling constant. The electron operators $c_{n,i,\sigma}$ 
and $c_{n,i,\sigma}^\dag$ satisfy the periodic boundary conditions for
both the leg and the rung directions,
$$c_{n+L,i,\sigma}
= c_{n,i,\sigma}, \ \ \ c_{n,i+3,\sigma}=c_{n,i,\sigma},
$$ 
and anticommutation relations, 
\begin{eqnarray*}
&&\{c_{m,i,\sigma},c_{n,j,\tau}^\dag \}=\delta_{m,n} \delta_{i,j}
 \delta_{\sigma,\tau}, \nonumber \\ 
&&\{c_{m,i,\sigma},c_{n,j,\tau} \}=0, 
\ \ \ \{c_{m,i,\sigma}^\dag, c_{n,j,\tau}^\dag \}=0.
\end{eqnarray*}
The hopping parameters are 
given by $t>0$, $s_{1,2}=s_{2,3}=s>0$, and $s_{3,1}= \beta s > 0$. 
The strong coupling expansion shows that
this model at the half-filling case is reduced to 
the Heisenberg model with $J_1=4t^2/U$, $J_{\rm r}= 4s^2/U$, and 
$\alpha = \beta^2$.

By performing the suitable unitary transformation, 
the hopping Hamiltonian can be mapped to the following diagonal form:
\begin{equation}
H_{\rm hop} = \sum_{k} \sum_{i=1}^3 \sum_{\sigma = \uparrow, \downarrow} 
E_{i}(k)d_{k,i,\sigma}^\dag d_{k,i,\sigma},
\end{equation}
where the wave number $k$ is summed over $\frac{2 \pi}{L} \leq k \leq 2 \pi$.
The operators $d_{k,i,\sigma}$ and $d_{k,i,\sigma}^\dag$ are defined by
\begin{subequations}
\begin{eqnarray}
d_{k,i,\sigma} = 
\frac{1}{\sqrt{L}}\sum_{n=1}^L \sum_{j=1}^3 e^{-\imath kn} O_{ij} c_{n,j,\sigma},  \\
d_{k,i,\sigma}^\dag= 
\frac{1}{\sqrt{L}}\sum_{n=1}^L \sum_{j=1}^3 e^{\imath kn} O_{ij} c_{n,j,\sigma}^\dag,
\end{eqnarray}  
\end{subequations}
which satisfy the standard anticommutation relations
\begin{eqnarray}
&&\{d_{k,i,\sigma},d_{l,j,\tau}^\dag \}=
\delta_{k,l} \delta_{i,j} \delta_{\sigma,\tau}, \nonumber \\ 
&&\{d_{k,i,\sigma},d_{l,j,\tau} \}=0, 
\ \ \ \{d_{k,i,\sigma}^\dag, d_{l,j,\tau}^\dag \}=0.
\end{eqnarray} 
The explicit form of the orthogonal matrix is
\[
O=\left(
\begin{array}{ccc}
 -\frac{1}{\sqrt{2}} & 0 & \frac{1}{\sqrt{2}} \\
  u_+ & v_+ & u_+ \\
  u_- & v_- & u_-
\end{array}
\right),
\]
where the matrix elements are given by
\begin{eqnarray}
&&u_\pm=\frac{1}{n_\pm}, \ \ \ \ \ v_\pm= \frac{c_\pm}{n_\pm}, 
\ \ \ \ \ n_\pm = \sqrt{2+c_\pm^2}, \nonumber \\ 
&&c_\pm = 
\frac{3 \beta \pm \sqrt{\beta^2+8}}{\beta^2+2 \pm \beta\sqrt{\beta^2+8}}.
\end{eqnarray}
The energy
eigenvalues of the one-electron states are
\begin{subequations}
\begin{eqnarray}
&&E_{1}(k)= -\beta s +2 t \cos k, \\
&&E_{2}(k)= \frac{1}{2}(\beta s - s\sqrt{\beta^2 + 8} + 4 t \cos k), \\
&&E_{3}(k)=  \frac{1}{2}(\beta s + s\sqrt{\beta^2 + 8} + 4 t \cos k).
\end{eqnarray}
\end{subequations}
Note that a degeneracy $E_1(k)=E_2(k)$ appears at $\beta=1$ due to the
translational ($\vec S_{i,j}\to \vec S_{i+1,j}$) and the parity 
($\vec S_{1,j}\leftrightarrow \vec S_{3,j}$) 
symmetries along the rung direction.

For the half-filled case, $3L$ one-electron 
states should be occupied by electrons with up and down spins.
As a result, the ground state of the hopping Hamiltonian 
has one, two or three pairs of the Fermi points $(k_j,\bar{k}_j)$
just on the Fermi sea, depending on the parameters $s/t$ and $\beta$. 
Since the low-energy excitations are
given by the particle-hole creations around these Fermi points,
they may be represented by using the Dirac fermions, the left mover 
$\psi_{j,\sigma}(x)$ and the right one $\bar{\psi}_{j,\sigma}(x)$, 
which are defined from the electrons around the $j$-th pair of Fermi points. 
If the $j$-th band has no Fermi points in the half-filled case, 
we should neglect $\psi_{j,\sigma}(x)$ and 
$\bar{\psi}_{j,\sigma}(x)$. On this understanding, 
we approximate the original electron operators in terms of
the Dirac fermions as follows: 
\begin{equation}
c_{n,i,\sigma} \sim \sqrt{a} \sum_{j=1}^3 O^{-1} _{ij}
(e^{\imath k_j x/a} \psi_{j,\sigma}(x)
+e^{\imath \bar{k}_j x/a}\bar{\psi}_{j,\sigma}(x)),
\end{equation} 
where $a$ is the lattice spacing with dimension of length and
$x=a n$ is the continuous position coordinate.

For this free Dirac fermion system, we take into account 
the effects of the on-site Coulomb interaction~(\ref{coulomb}) 
as the perturbation, and we use the non-Abelian bosonization
techniques. Following naively the field-theory argument in 
Ref.~\onlinecite{Aff-Hal}, 
one can expect that when the number of Fermi-point pairs is odd (even), 
the spin excitations are gapless (gapped) in the half-filled Hubbard
tube. In particular, in the cases of one or two Fermi-point pairs, 
we can explicitly determine whether or not a spin gap exists as follows.

First, we consider the case of one pair of Fermi points
$k_1=\frac{3\pi}{2}$ and $\bar{k}_1=\frac{\pi}{2}$. 
In this case, the interaction~(\ref{coulomb}) is approximated as the sum
of an Umklapp interaction and two marginal ones
$$
H_{\rm int} \sim \int dx \Big[g_1 \Theta_1(x)-g_2 \Theta_2(x)-g_3
\Theta_3(x)+\cdots\Big], 
$$
where $g_{1,2,3}$ are positive coupling constants proportional to $U$.
The Umklapp term is expressed as 
\begin{equation}
\label{Umklapp}
\Theta_1(x)=\psi_{1,\uparrow}(x)^\dag
\psi_{1,\downarrow}(x)^\dag \bar{\psi}_{1,\uparrow}(x)
\bar{\psi}_{1,\downarrow}(x),
\end{equation}
and the marginal interaction between the U(1) charge currents 
is given by
$$
\Theta_2(x) = 
\psi_{1}(x)^\dag  
\psi_{1}(x) \bar{\psi}_{1}(x)^\dag \bar{\psi}_{1}(x),
$$
where $\psi_{1}={}^t(\psi_{1,\uparrow},\psi_{1,\downarrow})$. 
It is known that the bosonized form of $\Theta_{1,2}$ contains only 
the charge degrees of freedom and they open a charge gap 
when $g_2$ is positive. 
Then, the remaining spin degrees of freedom 
are described by the gapless level-1 SU(2) WZW
theory.~\cite{affleck,tsvelik,gogolin} 
This phenomenon, i.e., the charge-spin separation 
is well-known in the single Hubbard chain model. 
For this WZW theory, the third interaction 
\begin{equation}
\label{current_int}
\Theta_3(x) =
\psi_{1}(x)^\dag \frac{\vec \sigma}{2} 
\psi_{1}(x) \cdot \bar{\psi}_{1}(x)^\dag 
\frac{\vec \sigma}{2}\bar{\psi}_{1}(x),
\end{equation}
is known to be marginally irrelevant if $g_3>0$. 
The coupling constant $g_3$ 
is hence renormalized to be zero in the low-energy limit. 
Except for the above interactions $\Theta_{1,2,3}(x)$, 
there is no relevant operator with the invariance 
under the one-site translation along the leg,
\begin{equation}
\psi_{1,\sigma}(x) \rightarrow
e^{\imath k_1} \psi_{1,\sigma}(x), \ \ \  
\bar{\psi}_{1,\sigma}(x) \rightarrow 
e^{\imath \bar{k}_1} \bar{\psi}_{1,\sigma}(x), 
\end{equation}
as in the case of the Heisenberg chain. 
The spin excitations, therefore, remain gapless. 

On the other hand, when there exist two pairs of the 
Fermi points, $(k_1,\bar{k}_1)$ and $(k_2,\bar{k}_2)$, the fate of 
the spin excitations is different from the above scenario. 
In this case, the spin sector in the hopping part of the Hubbard tube 
is described by a level-2 SU(2) WZW theory, which is derived from
two decoupled Dirac fermions.~\cite{Note1} 
The Coulomb interaction yields several perturbations for this
theory. For example, 
applying the non-Abelian bosonization rule,~\cite{affleck} 
we find that an interaction, derived from Eq.~(\ref{coulomb}),  
\begin{equation}
\label{relevant_level2}
\psi_{1,\uparrow} (x)^\dag\bar{\psi}_{2,\uparrow}(x)
\bar{\psi}_{1,\downarrow}(x)^\dag \psi_{2,\downarrow}(x),
\end{equation}
contains a relevant perturbation for the level-2 WZW model, and 
it is invariant under the translational operation 
\begin{equation}
\psi_{j ,\sigma}(x) \rightarrow
e^{\imath k_j} \psi_{j, \sigma}(x), \ \  \bar{\psi}_{j, \sigma}(x) \rightarrow
e^{\imath \bar{k}_j} \bar{\psi}_{j, \sigma}(x).
\end{equation}
Particularly for $\beta=1$, the number of possible 
relevant operators with the translational invariance is increased,
because of the coincident Fermi points $k_1=k_2$ and $\bar{k}_1=\bar{k}_2$.
Therefore, we conclude that 
any gapless spin excitation generally has no chance to survive 
except particularly rare cases [e.g., when the coupling constant of
Eq.~(\ref{relevant_level2}) is zero].

\begin{figure}[htbp]
\begin{center}
\includegraphics[width=0.8\linewidth]{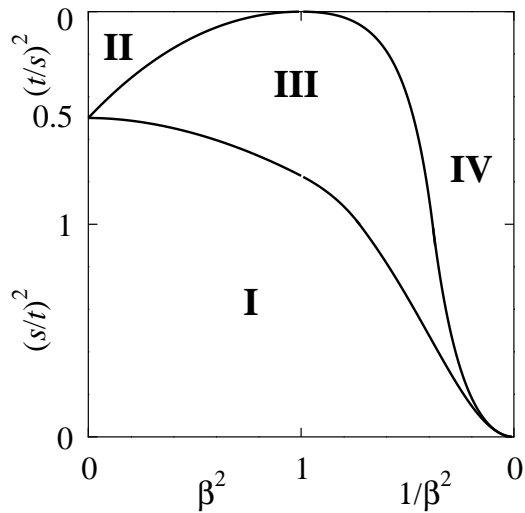}%
\caption{Phase diagram obtained from 
the effective Hubbard model~(\ref{Hubbard_tube}). 
In the strong-coupling limit ($U/t,U/s\gg 1$), the horizontal axis 
$\beta^2$ and vertical one $(s/t)^2$ can be regarded as 
$\alpha$ and $J_{\rm r}/J_1$, respectively. 
The effective theory claims that the phases (II) and (IV) are 
gapless and the phase (III) is gapful. The phase (I) is possibly gapless.}
\label{phase_eff}
\end{center}
\end{figure}

Since for the case of three Fermi-point pairs, 
the interactions among three Dirac fermions, generated from
Eq.~(\ref{coulomb}), are fairly complicated, 
it is difficult to analyze them and judge whether or not the spin 
excitation can survive as gapless. 
However, as we already stated, it can be expected from
Ref.~\onlinecite{Aff-Hal} that the spin excitation 
is presumably gapless in this case of three Fermi-point pairs.

From these arguments, we can draw the ground-state phase diagram of
the half-filled Hubbard tube as shown in  Fig.~\ref{phase_eff}.
The phase (I) has three pairs of the Fermi points,
the phases (II) and (IV) have one pair, and the central phase (III) has two pairs.
Therefore, we can predict that the phases (II) and (IV) have gapless spin
excitations, whereas the phase (III) possesses a spin gap. 
If $(s/t)^2$ and $\beta^2$ are, respectively, 
replaced with the coupling ratio $J_1/J_{\rm r}$ and the
asymmetric parameter $\alpha$ in Fig.~\ref{phase_eff}, 
one may interpret the phase diagram as
that of the $S=\frac{1}{2}$ three-leg spin tube~(\ref{ham}).
Although we have treated the on-site interaction~(\ref{coulomb}) 
perturbatively, we have assumed that the weakly-interacting case
($U/t,U/s\ll 1$) is smoothly linked to the strongly-interacting one, 
which is nothing but the spin tube~(\ref{ham}).

We here note that the gapful phase (III) is predicted to be extended 
around the line $\beta=1$ for a finite $s/t$. In the limit 
$s/t \rightarrow\infty$, the both the left- and right-side phase
boundaries of the region (III) converge to $\beta=1$. 
This narrowing of the phase (III) is consistent with the numerical
results~\cite{sakai,arikawa} 
in the strong-rung-coupling limit 
$J_{\rm r}/J_1 \rightarrow \infty$. 
We will discuss this limit in more detail in Sec.~\ref{energy_gap}.

Finally, we briefly argue the universality classes 
of the phase transitions at two phase boundaries, 
(II)-(III) and (III)-(IV). 
For the level-1 SU(2) WZW model in the phases (II) and (IV), 
the most relevant perturbation is 
the marginal current-current interaction,~\cite{affleck,tsvelik,gogolin}
Eq.~(\ref{current_int}),  
which is invariant under the translational and spin-rotational
operations. 
Since it possibly becomes marginally ``relevant'' when parameters are
finely tuned and then $g_3$ becomes negative, 
we speculate that the transition from the phases (II) or (IV) 
to (III) is caused by this marginal term. Therefore, the transitions are
expected to be in the BKT universality
class. This speculation may be naturally accepted if we recall 
the following two known results of 
the $S=\frac{1}{2}$ zigzag Heisenberg chain, namely, 
the spin chain with the nearest- and the next-nearest-neighbor
interactions:~\cite{on} (i) When the next-nearest-neighbor hopping is
sufficiently small in the half-filled electron system on zigzag lattice, 
one obtains one pair of Fermi points. The spin
excitations are therefore described by a level-1 SU(2) WZW model, like
the phases (II) and (IV). 
(ii) It has been numerically shown~\cite{on} that when 
the next-nearest-neighbor interaction is increased in the zigzag spin
chain, the BKT 
transition takes place and the ground state changes 
into a dimerized state from a Tomonaga-Luttinger
liquid described by a level-1 SU(2) WZW model.

On the basis of this speculation, we numerically analyze the phase
transitions between (II)-(III) and (III)-(IV) in Sec.~\ref{numerical}.

\subsection{Gapless phase for $J_{\rm r} \ll J'_{\rm r} \ll J_1 $}
\label{gapless_phase}
In Sec.~\ref{global}, we have obtained a qualitative 
phase diagram of the spin tube~(\ref{ham}) as shown in 
Fig.~\ref{phase_eff}. However, some subtle points still remains, and
particularly it is doubtful whether or not there is the gapless phase (I).   
To partially resolve these issues, we here focus on the extreme situation 
$J_{\rm r} \ll J'_{\rm r} \ll J_1, \ (\alpha =J'_r/J_{\rm r}\gg 1)$, 
which corresponds to the right-lower regime in Fig.~\ref{phase_eff}.
In this regime, we prove the existence of the gapless phase (I)
and discuss the phase transition between the gapless phase (I) and 
the gapful phase (III).

Before embarking on our analysis, 
we sketch the scenario of this subsection. 
We introduce three level-1 SU(2) WZW theories 
for the three decoupled Heisenberg chains,~\cite{affleck,tsvelik,gogolin} 
and then we treat their rung couplings as the perturbation, because 
the weak-rung-coupling regime $J_{\rm r}, J'_{\rm r} \ll J_1$ is
considered now. 
The first and the third chains are coupled to each other 
with $J'_{\rm r}$ which is much stronger than 
two remaining couplings $J_{\rm r}$. 
It is well-known that the $J'_{\rm r}$ coupling involves a relevant 
interaction with conformal dimensions $(\frac{1}{2},\frac{1}{2})$ 
in the two coupled WZW models [see Eq.~(\ref{mass})], 
and it produces an energy gap.~\cite{shelton} 
On the other hand, like the case of one Fermi-point pair in
Sec.~\ref{global}, the WZW model for the second chain 
has the marginal irrelevant current-current interaction,
\begin{equation}
\label{marginal}
\lambda_2 \Phi_2, 
\end{equation}
with a finite negative coupling constant $\lambda_2 < 0$. 
The operator $\Phi_2$ is equivalent to 
Eq.~(\ref{current_int}), if we use the Dirac fermions 
$(\psi_{1,\sigma},\bar\psi_{1,\sigma})$ to describe the second chain. 
The negative sign makes Eq.~(\ref{marginal}) irrelevant and the second
chain is gapless. A weak rung coupling $J_{\rm r}$ between this WZW model and 
the massive theory for the two coupled chains must give a correction to 
the coupling constant $\lambda_2$. 
If $J_{\rm r}$ is sufficiently small, the sign of 
$\lambda_2 < 0$ would not change and the gapless excitation is
preserved. These arguments convince us 
that the gapless phase (I) definitely exists. 
Furthermore, we expect the phase transition from the gapless phase (I) 
to the gapful phase (III). Namely, if $J_{\rm r}$ exceeds a critical
value, the coupling constant $\lambda_2$ might change to be positive. 
In this case, the marginal operator (\ref{marginal}) 
becomes relevant, which produces an excitation gap.
This transition between the phases (I) and (III) 
is nothing but of the BKT 
type,~\cite{bere,kt} 
as in the zigzag Heisenberg chain.~\cite{on}

Now, let us calculate the correction to $\lambda_2$ in an explicit way
to confirm the above sketch.
We represent the partition function of the three coupled 
WZW models through the path integral formalism. Note that the SU(2)
spin-rotational symmetry is conserved automatically in the language of
the WZW models, i.e., the non-Abelian bosonization. 
The three WZW models can be 
represented in six Dirac fermions ${\Psi}_{i,\alpha}(z,\bar{z}) 
\,\, [i=1,2,3, \,\rm{and}\,\, \alpha= \uparrow, \downarrow]$ 
and three ghost bosons $\phi_i(z,\bar{z}) \,\, [i=1,2,3]$. Here,
$z=v\tau+\imath x$ and $\bar z=v\tau-\imath x$ are the chiral
coordinates ($\tau$: imaginary time) and 
the index $i$ denotes the chain number of the spin tube~(\ref{ham}). 
We employ this free-field representation.~\cite{polyakov,IM,IK} 
The Lagrangian density for the decoupled three chains is given by
\begin{equation}
{\cal L}_0(z,\bar{z})=2 \sum_{i=1}^3 \left[
\sum_{\alpha=\uparrow,\downarrow}
(\bar{\Psi}_{i,\alpha} ^\dag \partial \bar{\Psi}_{i,\alpha}
+ \Psi_{i,\alpha}^\dag  \bar{\partial} \Psi_{i,\alpha})
-\partial \phi_i \bar{\partial} \phi_i \right],
\label{L0}
\end{equation}
where we have decomposed the Dirac fermions into the linear combination of
the left mover $\Psi_{i,\alpha}(z)$ and the right mover 
$\bar{\Psi}_{i,\alpha}(\bar{z})$ 
as $\Psi_{i,\alpha}(z,\bar{z})=\Psi_{i,\alpha}(z)
+\bar{\Psi}_{i,\alpha}(\bar{z})$.~\cite{Note3} 
The nonvanishing two-point correlation functions of these fields, 
calculated from this free Lagrangian (\ref{L0}), are
\begin{subequations} 
\begin{eqnarray}
\langle \Psi_{i,\alpha}(z)^\dag \Psi_{j,\beta}(w) \rangle
&=& \frac{\delta_{ij}\delta_{\alpha \beta}}{z-w},
\label{freea} 
\\
\langle \bar{\Psi}_{i,\alpha}(\bar{z}) ^\dag \bar{\Psi}_{j,\beta}(\bar{w}) 
\rangle &=&\frac{\delta_{ij}\delta_{\alpha \beta}}{\bar{z}-\bar{w}}, 
\label{freeb} 
\\ \label{freec} 
\langle e^{\imath \phi_i(z,\bar{z})} e^{-\imath \phi_i(w,\bar{w})} 
\rangle &=&\delta_{ij}|z-w|.
\end{eqnarray}
\end{subequations} 
The anomalous correlation of the ghost bosons should be noted. 
These ghost fields
kill the charge degrees of freedom and extract the 
gapless spin degrees of freedom in the Dirac fermions. 
The primary field $g_{\alpha \beta}^i(z,\bar{z})$ with the conformal 
dimension $(\frac{1}{4},\frac{1}{4})$
in the $i$-th level-1 SU(2) WZW theory
is represented in terms  of these free fields:  
$$
g_{\alpha \beta}^i(z,\bar{z})= 
\Psi_{i,\alpha} (z)^\dag\bar{\Psi}_{i,\beta}(\bar{z}) e^{\imath \phi_i(z,\bar{z})}.
$$
Furthermore, the spin operators are represented as
$$
\vec S_{i,n}/a \approx \vec J_i(z,\bar z)+(-1)^n \vec N_i(z,\bar z),
$$
in which the smooth and the staggered parts are given by 
\begin{subequations} 
\begin{eqnarray}
\vec J_i(z,\bar z) &=& \Psi_i ^\dag \frac{\vec{\sigma}}{2}\Psi_i+ 
\bar{\Psi}_i ^\dag \frac{\vec{\sigma}}{2} \bar{\Psi}_i,\\
\vec N_i(z,\bar z) &=& C_0 
\left( \Psi_i ^\dag \frac{\vec{\sigma}}{2} \bar{\Psi}_i
 e^{\imath \phi_i} +{\rm h.c.}
\right),
\label{jn}
\end{eqnarray}
\end{subequations} 
with $\Psi_i={}^t(\Psi_{i,\uparrow},\Psi_{i,\downarrow})$. Here $C_0$ is
a nonuniversal constant. Applying Eqs.~(\ref{freea})-(\ref{freec}), 
one can evaluate the asymptotic forms of several correlation functions 
for the field $g_{\alpha \beta}^i$ and the spins $\vec S_{i,n}$. 
Among the spin-rotational- and the translational-symmetric operators, 
the most relevant coupling between the $i$-th and the $j$-th WZW models
is 
\begin{eqnarray}
\label{mass}
\Phi_{ij}(z,\bar{z}) &= & 
\vec N_i(z,\bar z)\cdot\vec N_j(z,\bar z),
\end{eqnarray}
with the conformal dimensions $(\frac{1}{2},\frac{1}{2})$. As we already
mentioned, this relevant term $\Phi_{31}(z,\bar{z})$ produces the energy gap
in the two strongly coupled WZW models. The WZW model for the second
chain interacts with these coupled WZW models 
via the weak rung coupling $\Phi_{12}(z,\bar{z})$ and $\Phi_{23}(z,\bar{z})$. 
The explicit form of the marginal operator (\ref{marginal}) 
of the second WZW model is
\begin{equation}
\Phi_2(z,\bar{z})= \Psi^\dag_2 \frac{\vec{\sigma}}{2} \Psi_2 
\cdot \bar{\Psi}_2^\dag\frac{\vec{\sigma}}{2} 
\bar{\Psi}_2.
\label{marginaldef}
\end{equation}

From these materials, the total Lagrangian density for the asymmetric 
spin tube under the condition $J_{\rm r} \ll J'_{\rm r} \ll J_1$ 
is written as  
\begin{eqnarray}
{\cal L}(z,\bar{z}) &=& {\cal L}_0(z,\bar{z})  
+ 
\lambda_{31} \Phi_{31}(z, \bar{z}) 
+\lambda_{12} \Phi_{12}(z,\bar{z}) \nonumber\\
&&+ \lambda_{23} \Phi_{23}(z, \bar{z})
+ \lambda_2 \Phi_{2}(z,\bar{z})+\cdots,
\end{eqnarray}
where the coupling constants are 
$\lambda_{31} \propto J'_{\rm r}$, 
$\lambda_{12}=\lambda_{23}\propto J_{\rm r}$, and $\lambda_2 < 0$. 
Therefore, the partition function is 
written in the path integral over Grassmann and
boson fields as 
\begin{eqnarray*}
Z &=& \int \prod_{i,\alpha}{\cal D} \bar{\Psi}_{i,\alpha}
{\cal D} \bar{\Psi}_{i,\alpha} ^\dag 
{\cal D} \Psi_{i,\alpha}  {\cal D}\Psi_{i,\alpha}^\dag 
{\cal D}\phi_i \\
&&\times\exp \left[-\int \frac{d^2z}{2\pi}{\cal L}(z,\bar{z}) \right].
\end{eqnarray*}
Under the condition $J_{\rm r} \ll J'_{\rm r} \ll J_1$, the low-energy
physics must be governed by the second chain weakly coupled to the other
two chains. To obtain its effective theory, 
we may integrate out the massive degrees of freedom 
$\Psi_{i,\alpha}(z,\bar{z})$ and $\phi_i(z,\bar{z})$ with $i=1,3$.
To calculate the correction to the coupling constant $\lambda_2$,
we regard the relevant term $\Phi_{31}(z, \bar{z})$ as an unperturbed
Lagrangian and expand the partition function in all the other operators 
${\cal L}_0$, $\Phi_{12}(z, \bar{z})$, $\Phi_{23}(z, \bar{z}), \cdots$. 
This expansion can be performed by a lattice regularization 
of two-dimensional Euclidean spacetime. In this regularization,
the partition function is represented as the following 
multiple integration over countable variables:
\begin{eqnarray}
\label{partition}
Z &=& \int \prod_{n \in N}  \prod_{i \in I} 
\prod_{\alpha \in S} 
\Big\{ a^2 d \bar{\Psi}_{i,\alpha}(\bar{z}_{n} ) 
d \bar{\Psi}_{i,\alpha} ^\dag (\bar{z}_{n}) \\
&&d \Psi_{i,\alpha} (z_n) 
d \Psi_{i,\alpha}^\dag (z_n) 
d \phi_i(z_n,\bar{z}_{n})
{\exp} 
\left[-\frac{a^2}{2 \pi}{\cal L}(z_n, \bar{z}_n) 
\right] 
\Big\}, \nonumber
\end{eqnarray}
where
$I=\{1,2,3\}$, $S=\{\uparrow, \downarrow\}$, 
and $N=\{(n_1,n_2) \in \Z^2| 1 \leq  n_1, n_2 \leq L \}$
is a finite set of integer pairs with a large number of elements.
For an integer pair 
$n=(n_1, n_2) \in N$, we define discretized coordinates
$(z_{n},\bar{z}_n)=a(n_1+\imath n_2,n_1-\imath n_2)$
for $(z, \bar{z})$  with
a lattice spacing parameter $a$. 
For a finite number of Grassmann variables,
the Taylor expansion is reduced to a finite summation,
$$
{\exp} \left[-\frac{a^2}{2 \pi}{\cal L}(z_n, \bar{z}_n) \right]
= \sum_{k=1} ^{K} 
\frac{1}{k !}\left[-\frac{a^2}{2 \pi}{\cal L}(z_n, \bar{z}_n) \right]^k,
$$
with a certain positive integer $K \leq 12|N|$, 
due to the nilpotency of the Grassmann variables, 
\begin{eqnarray*}
\Psi_{i,\alpha}(z_n)^2=0, \ \  \bar{\Psi}_{i,\alpha}(\bar{z}_n) ^2=0,\\
\Psi_{i,\alpha}^\dag(z_n)^2=0, \ \ \bar{\Psi}_{i,\alpha}^\dag(\bar{z}_n) ^2=0.
\end{eqnarray*}
The following integration formula of Grassmann variables,
\begin{eqnarray}
&&\int \prod_{n \in N} \prod_{i \in I} 
\prod_{\alpha \in S} \left( a^2
d \bar{\Psi}_{i,\alpha}(\bar{z}_{n} ) 
d \bar{\Psi}_{i,\alpha}^\dag (\bar{z}_{n}) 
d \Psi_{i,\alpha} (z_n) 
d \Psi_{i,\alpha}^\dag (z_n)
\right)  \nonumber 
\\ && \times 
\prod_{n \in M} \prod_{i \in J} 
\prod_{\alpha \in T}\left( a^2\Psi_{i,\alpha}^\dag (z_n)\Psi_{i,\alpha} (z_n)
\bar{\Psi}_{i,\alpha}^\dag(\bar{z}_{n}) \bar{\Psi}_{i,\alpha}(\bar{z}_{n} ) 
\right) 
\nonumber 
\\ &&
=\left\{
\begin{array}{ccl}
1 & : & M=N, J=I, T=S   
\\
0 & : & {\rm otherwise,} 
\end{array}
\right. \label{integration}
\end{eqnarray}
is important to calculate this expansion.
In addition, using an equality for two Grassmann variables
$$
\Psi_{\alpha} \Psi_\beta = \imath \sigma^2 _{\alpha \beta} 
\Psi_\uparrow \Psi_\downarrow,\,\,\,(\sigma^{1,2,3}=\sigma^{x,y,z})
$$
and a trace formula of the Pauli matrices
$$
{\rm tr}[\sigma^a \sigma^2~^t\sigma^b \sigma^2]
= \sum_{\alpha,\beta,\gamma,\delta} \sigma^a_{\alpha\beta} 
\sigma^2_{\beta\gamma} \sigma^b _{\delta\gamma}
\sigma^2_{\delta\alpha}=-2 \delta^{a b},
$$
one can reduce products of Pauli matrices and 
Grassmann variables to the following single term,  
\begin{eqnarray}
\label{grass}
&&\prod_i \Psi^\dag_i \frac{{\sigma}^a}{2} 
\bar{\Psi}_i \ \Psi_i ^\dag\frac{{\sigma}^b}{2} 
\bar{\Psi}_i \ \Psi^\dag_i \frac{{\sigma}^c}{2} 
\Psi_i \ \bar{\Psi}_i ^\dag\frac{{\sigma}^d}{2} 
\Psi_i \nonumber \\
&&= \prod_i \Psi_{i,\uparrow}^\dag 
\Psi_{i,\downarrow}^\dag \Psi_{i,\uparrow} 
\Psi_{i,\downarrow} \bar{\Psi}_{i,\uparrow}^\dag 
\bar{\Psi}_{i,\downarrow}^\dag
\bar{\Psi}_{i,\uparrow} \bar{\Psi}_{i,\downarrow} 
\frac{\delta^{ab}}{2} \frac{\delta^{cd}}{2}.\,\,\,\,\,\,\,\,\,\,\,\,
\end{eqnarray}
These formulas (\ref{integration}) and (\ref{grass}) 
make it easier to calculate the expansion of the partition function 
(\ref{partition}). As a result, we obtain 
\begin{eqnarray}
\label{Z_final}
&&Z 
=\int \prod_{n \in N}  
\prod_{\alpha \in S}[a^2
d \bar{\Psi}_{2,\alpha}(\bar{z}_{n} ) 
d \bar{\Psi}_{2,\alpha}^\dag (\bar{z}_{n}) \nonumber \\ 
&&d \Psi_{2,\alpha} (z_n) d \Psi_{2,\alpha}^\dag (z_n) 
d \phi_2(z_n,\bar{z}_{n})] \nonumber \\  
&&\times\prod_{n \in N}\Big\{ 
{\exp} \Big[-\frac{a^2 \lambda_2}{2 \pi}\Phi_2(z_n, \bar{z}_n) + \cdots \Big]  
\nonumber \\
&& \Big[\frac{15}{32}\frac{\lambda_{31}^4}{(2 \pi)^4}-\frac{5}{16} 
\frac{\lambda_{31}^3 \lambda_{12} \lambda_{23}}{(2 \pi)^5}
\Phi_2(z_n, \bar{z}_n) + \cdots \Big] \Big\} \nonumber \\
&&\sim
\int \prod_{\alpha}{\cal D} \bar{\Psi}_{2,\alpha}
{\cal D} \bar{\Psi}_{2,\alpha} ^\dag 
{\cal D} \Psi_{2,\alpha}  {\cal D} \Psi_{2,\alpha} ^\dag {\cal D}
\phi_2  \nonumber \\
&&{\exp} \left[-\int \frac{d^2z }{2 \pi}\left(\lambda_2+ 
\frac{2 \lambda_{12}\lambda_{23}}{3 \lambda_{31}}\right)
\Phi_2(z, \bar{z}) + \cdots \right]. \ \ \ \ \  
\end{eqnarray}
Here, we have neglected several terms of charge degrees of freedom.
The final expression in Eq.~(\ref{Z_final}) clearly indicates 
that the correction to $\lambda_2$ is 
$2\lambda_{12}\lambda_{23}/(3 \lambda_{31})=2J_{\rm r}/(3 \alpha)$, 
and the phase boundary between (I) and (III) is given by
\begin{equation}
\label{boundary}
J_{\rm r}=- \frac{3}{2}\lambda_2  \alpha. \,\,\,(\lambda_2<0).
\end{equation}
Thus, the gapless excitation in the second
chain is preserved and the gapless phase is expanded under 
the condition $J_{\rm r} \ll J'_{\rm r} \ll J_1$. Namely, we have proved
that the phase (I) is exactly present at least in the region 
$J_{\rm r} \ll J'_{\rm r} \ll J_1$. Furthermore, we find that 
if $J_{\rm r}$ is large enough to change 
the sign of $\lambda_2+ \frac{2 J_{\rm r}}{3 \alpha}$, 
$\Phi_2(z,\bar{z})$ becomes relevant and an energy gap appears. 
Since this phase transition is induced by the marginal operator 
$\Phi_2(z,\bar{z})$, it belongs to the BKT 
universality class, as in the 
$S=\frac{1}{2}$ zigzag Heisenberg chain. This gapped state must
correspond to the phase (III) in Fig.~\ref{phase_eff}. 
In the gapful region under $J_{\rm r} \ll J'_{\rm r} \ll J_1$, 
the second chain must be dimerized. On the other hand, 
as we have often mentioned, 
the gapful phase in the vicinity of the symmetric line $\alpha=1$
has the valence-bond order. Since both the orders break the same
translational symmetry, the state with the dimerized second chain would
smoothly change into the valence-bond ordered state when
we vary $\alpha$ from a large value to unity in the gapful phase (III).

Finally, we notice that in the prediction~(\ref{boundary}), 
the gapless phase (I) becomes wider for larger $\alpha$, contrary 
to the phase diagram, Fig.~\ref{phase_eff}.  
Two reasons for this contradiction are immediately found. 
First, we should not precisely trust the location of 
the phase boundaries in Fig.~\ref{phase_eff}, which are 
drawn just by counting the number of Fermi points. 
The argument in Sec.~\ref{global} 
is only qualitatively correct to determine the phase boundaries.
Particularly, since the gapful phase (III) sandwiched 
by two gapless phases (I) and (IV) is quite
narrow for large $\alpha$ in Fig.~\ref{phase_eff}, then 
we cannot claim the existence of the gapful phase (III)
for $\alpha \gg 1$ by using Fig.~\ref{phase_eff}. 
Therefore, the phases (I) and (IV) might be smoothly connected 
in the large-$\alpha$ region. 
For such a case, the transition between two regions (III) and (IV) 
is expected to be also of the BKT type. 
Second, the three Dirac fermions in the region (I) 
do not always imply the gapless phase,~\cite{Arrigoni,lin} 
as discussed in Sec.~\ref{global}. The coupled 
three Dirac fermions possibly produce an energy gap 
by the strong frustration, if $\alpha$ approaches 
the symmetric point $\alpha=1$. 
The result~(\ref{boundary}) also shows this tendency.

\subsection{Energy gap for $J_{\rm r} \gg J_1$ and 
$J_{\rm r} \sim J_{\rm r}'$} 
\label{energy_gap}
In this subsection, we 
discuss the existence of an energy gap 
for a sufficiently strong rung coupling $J_{\rm r} \gg J_1$ and 
a sufficiently weak asymmetry $|\alpha-1| \ll 1$ except
at $J_1=0$. Namely, we explain that the phase (III) has a finite width 
along the line $J_{\rm r}/J_1={\rm const}\gg 1$.

As we mentioned before, Kawano and Takahashi~\cite{kawano} have obtained 
the spin-orbital type model 
as an effective theory for the symmetric ($\alpha=1$) spin tube 
with a strong rung coupling. The Hamiltonian is written as 
\begin{equation}
H_{\rm eff} ^{\rm sym} = 
\frac{J_1}{3 J_{\rm r}} \sum_{j=1}^L \vec{S}_j \cdot \vec{S}_{j+1} 
\Big[1+4 (\tau_j^+ \tau_{j+1} ^- + \tau_j ^-\tau_{j+1}^+)\Big],
\label{kawanohamil}
\end{equation}
which is the result of the first-order expansion in $J_1/J_{\rm r}$. 
The orbital Pauli matrices $\tau_i^\nu$ $(\nu = x, y, z)$
are defined for the states with respect to the left- and right-
handed spin configurations on each rung (see
Refs.~\onlinecite{kawano} and \onlinecite{arikawa}).
From this theory~(\ref{kawanohamil}), 
Kawano and Takahashi have shown that the ground state has a valence-bond
order with the translational symmetry $(\vec S_{i,j}\to\vec S_{i,j+1})$ 
spontaneously broken, and a finite excitation gap exists. 
It is believed that the energy gap is generated by the strong 
coupling between the spin and orbital degrees of freedom
in the effective Hamiltonian (\ref{kawanohamil}). Obviously,
either of them becomes gapless, if the other is frozen.

Quite recently, Nishimoto and Arikawa~\cite{arikawa} have extended 
the effective theory to the asymmetric case as follows:
\begin{eqnarray}
&&H_{\rm eff}^{\rm asym}
= H_{\rm eff} ^{\rm sym}+ (\alpha-1 ) \sum_{j=1}^L \tau_j^x. 
\label{arikawahamil}
\end{eqnarray}
For simplicity, we have re-defined the orbital 
Pauli matrices $\tau_i^\nu$ through a unitary transformation.
Remarkably, the asymmetry induces a transverse external field 
coupled to the orbital spins. 
A sufficiently strong asymmetry $|\alpha -1| \gg 1$ hence 
yields the saturated polarization of the orbital spins. 
In this case, the effective Hamiltonian~(\ref{arikawahamil}) 
is reduced to that of the antiferromagnetic Heisenberg chain.
The spin degree thus survives as a gapless excitation. 
Next, let us focus on the region around the symmetric line $\alpha \sim 1$.
In the limit $J_1/J_{\rm r} \rightarrow 0$, 
the ground state has a saturated orbital 
spin at any asymmetric point $\alpha \neq 1$. 
For a finite $J_1/J_{\rm r}$, however, 
we expect an extended gapful phase around $\alpha=1$, if the energy gap 
exists at $\alpha=1$ due to the coupling 
between the spin and the orbital degrees of freedom. 
The finite energy gap does not vanish by an
infinitesimal external field $\alpha-1$. 
In other words, the magnetization process 
of the orbital spins should show a zero-magnetization plateau. 
Therefore 
an energy gap would also be present for sufficiently weak asymmetry
$|\alpha-1| \ll 1$. 

From these arguments, we find that the gapful phase surely exists 
in a finite region under the conditions $J_{\rm r} \gg J_1$ and 
$J_{\rm r} \sim J_{\rm r}'$, and its width becomes smaller with
decreasing $J_1/J_{\rm r}$. 
This is consistent with the result obtained in Sec.~\ref{global}.
\\

\section{Numerical Analysis}
\label{numerical}
In this section, we numerically analyze the asymmetric spin
tube~(\ref{ham}), taking into account the results in the preceding
Sec.~\ref{effective}. 
We show that the arguments based on the effective theories are in good
agreement with the results of the numerical calculations. 
We will use results of the numerical diagonalization 
up to $L=10$ for the periodic boundary system, 
and those of the DMRG up to $L=128$ for the open boundary system.

\subsection{Phenomenological renormalization}
\label{PRG}
\begin{figure}
\includegraphics[width=0.9\linewidth]{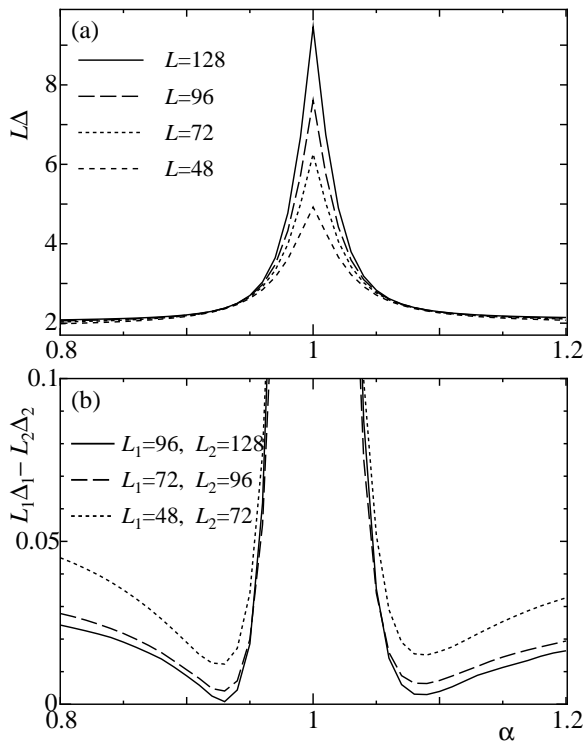}%
\caption{\label{pr} 
(a) Scaled gap calculated by DMRG for $J_1=0.5$. 
(b) Difference of the scaled gaps between two systems with 
sizes $L_1$ and $L_2$.}
\end{figure}

A useful order parameter to determine the phase boundaries  between the
gapless and the gapful phases in Fig.~\ref{phase_eff} 
is the spin gap $\Delta$, which is the energy gap between the singlet
ground state and the triplet excited state in the finite, but large system. 
We calculate it by means of the DMRG up to $L=128$, where we do not see
such a significant open boundary effect as local edge excitation. 
In actual DMRG computation, the number of retained bases is up to
$m=300$, within which well convergence is achieved for the scaled gap. 
In Fig.~\ref{pr} (a), the scaled gap $L\Delta$ is plotted versus
$\alpha$ for $J_1=0.5$. It indicates that the spin gap is just open in a
tiny region $\alpha \sim 1$ and it rapidly vanishes away from
$\alpha=1$. 
Thus we find two critical points $\alpha_{c1}$ and $\alpha_{c2}$ 
($\alpha_{c1}<1<\alpha_{c2}$), which are expected to be 
of the BKT 
type because of the wide gapless regions outside the gapful phase, 
and the discussion in Sec.~\ref{effective}.

\begin{figure}
\includegraphics[width=0.9\linewidth,angle=0]{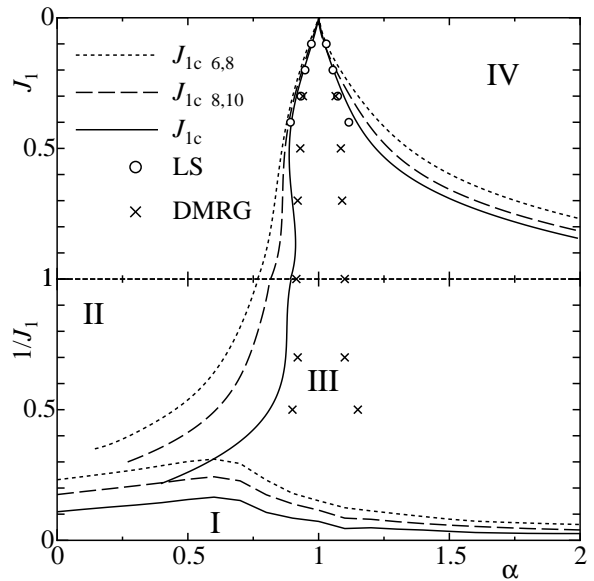}%
\caption{\label{phasen} 
Ground-state phase diagram of the asymmetric three-leg spin
 tube~(\ref{ham}), derived from the numerical analysis. 
The phases (I)-(IV) correspond to those in Fig.~\ref{phase_eff}. 
The cross points are determined by the DMRG. 
The three kinds of lines are done by the numerical diagonalization. 
The circle points are obtained from the level-spectroscopy method 
in Sec.~\ref{level_cross}. }
\end{figure}

In order to determine a phase boundary of the usual second order phase
transition, the phenomenological renormalization equation 
$L_1\Delta_{L_1}(\alpha_c)=L_2\Delta_{L_2}(\alpha_c)$ is often used
effectively. For the present critical point $\alpha_c$, however, 
this type of phenomenological renormalization has no clear crossing point. 
It seems from Fig.~\ref{pr} (a)  that the scaled gap $L\Delta$ increases 
with increasing $L$ in both gapless and gapful phases. 
This is because the scaled gap is an increasing function with respect to
$L$ not only in the gapful phase but also  in the gapless phase, since
the finite-size gap must have the logarithmic size correction term 
$\sim-1/\log L$. 
Here, we should recall that the logarithmic correction normally vanishes 
just at $\alpha_c$ due to the SU(2) symmetry in the $c=1$ CFT.~\cite{on} 
Therefore, instead of using the crossing point of the scaled gaps, we
can estimate $\alpha_c$ as a point where the size correction is
minimized. The difference of the scaled gap between two system 
sizes $L_1$ and $L_2$ is plotted versus $\alpha$ for $J_1=0.5$ in
Fig.~\ref{pr} (b). 
The minimum of the difference $L_1\Delta_{L_1}-L_2\Delta_{L_2}$ has a
very small $L$ dependence. 
Note that the minimum value $L_1\Delta_{L_1}-L_2\Delta_{L_2}$ decreases
as the size increases. This phenomenon and the assumption of the
BKT 
type transition suggest that this minimal value 
approaches zero as the system size increases. This is quite reasonable 
if we suppose the most important finite-size correction
to the scaled gap $L \Delta$ next to $1/ \log L$ term is order of 
$1/L^2$.\cite{on,cardy2,cardy3}
We thus determine $\alpha_c$ from the minima for two large systems
with $L_1=96$ and $L_2=128$ for $J_1<2$. 
The estimated $\alpha_{c1}$ and $\alpha_{c2}$ are shown as crosses in
Fig.~\ref{phasen}. 
They correspond to the phase boundary between two regions (II) and (III)
and that between (III) and (IV), respectively. 
At least these boundaries for the strong-rung-coupling regime $J_1\le 2$
are precise enough to justify that a finite gapful phase (III) exists. 
However, it is difficult to obtain $\alpha_c$ for $J_1 >2 $, because the
DMRG calculation is not well converged there.

In order to determine the phase boundaries for the weak-rung-coupling
regime $J_1>2$, 
we use the minimum points of $L_1\Delta_{L_1}-L_2\Delta_{L_2}$
calculated by the numerical diagonalization up to $L=10$ under the
periodic boundary condition. The estimated phase boundaries for 
$(L_1,L_2)=(6,8)$ and (8,10) are plotted as long-dashed and dashed
curves, respectively, in Fig.~\ref{phasen}. 
In addition, the infinite-$L$ curves extrapolated assuming the size
correction is proportional to $1/L^2$ in both directions of $J_1$ and
$\alpha$ are also shown as solid curves in Fig.~\ref{phasen}. 
At least the phase boundaries (II)-(III) and (III)-(IV) are consistent
with the DMRG results for $J_1 \ll 1$. 
The boundary (III)-(IV) is, however, significantly deviated from the
DMRG estimation for $J_1\sim 1 $. 
This discrepancy is supposed to be due to the error of extrapolation. 
This analysis also justifies the existence of the phase (I). 
However, the error of extrapolation becomes larger as we approach the
line $1/J_1=0$ in the case of $\alpha<1$. 
Thus it is difficult to conclude that the phase (I) really exists for
$\alpha<1$ within the present numerical demonstration. 
The boundary (I)-(III) will be discussed later. 
It is also difficult to confirm the boundary (I)-(II), 
and the phase (I) might combine with the phase (II) in a certain regime
with $\alpha<1$. 


\subsection{Conformal field theory analysis}
\label{CFT_scaling}
The numerical analysis based on CFT is generally efficient in
investigating the feature of quantum phase transitions 
in 1+1 dimensional systems. The conformal invariance gives the
$L$ dependences of the 
ground-state energy $E_0$ and the energy gaps for triplet excitations
$\Delta_{\rm t}$ and that for singlet excitations $\Delta_{\rm s}$
as the forms,
\begin{subequations}
\label{CFT_equations}
\begin{eqnarray}
{E_0 \over L} &\approx& \epsilon _{\infty} - {\pi v_s \over {L^2}} 
\frac{c}{6} + \cdots, \\
\Delta_{\rm t} &\approx& {\pi v_s \over L}\left(\eta - 
\frac{\sigma_{\rm t}}{\log L} 
+ \cdots\right), \label{triplet-gap} \\
\Delta_{\rm s} &\approx& {\pi v_s \over L}\left(\eta - 
\frac{\sigma_{\rm s}}{\log L} 
+ \cdots\right), \label{singlet-gap}
\end{eqnarray}
\end{subequations}
where $v_s$ is the spin wave velocity,
$c$ is the central charge, and $\eta$, $\sigma_{\rm t}$ and $\sigma_{\rm s}$ 
are the critical exponents. 
These exponents $\eta$ and $\sigma_{\rm t}$ 
appear in the spin correlation function~\cite{GS,IK}
$$\langle S_{i,0}^{\nu}S_{i,j}^{\nu}\rangle \sim (-1)^j 
(\log |j|)^{\sigma_{\rm t}} |j|^{-\eta},  \ \ \ (\nu =x,y,z.),  
$$
where $|j|\gg 1$. 
As discussed in Sec.~\ref{effective}, the gapless phases
(II) and (IV) are predicted to be described by the level-1 SU(2) WZW
model. This model indicates the universal constants $c=1$, 
$\eta=1$, $\sigma_{\rm t} = \frac{1}{2}$, and $\sigma_{\rm s} = -\frac{3}{2}$.

\begin{figure}
\includegraphics[width=0.85\linewidth,angle=0]{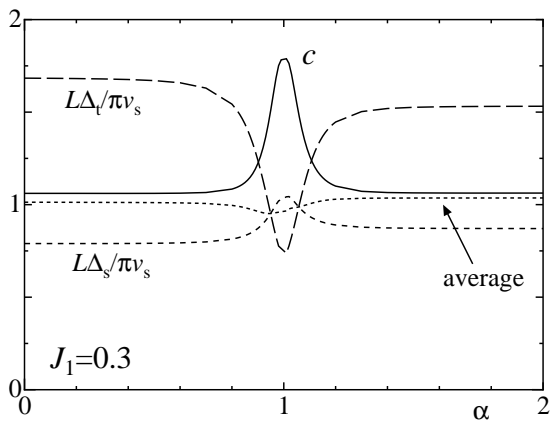}%
\caption{\label{c-eta} 
Central charge $c$ and critical exponents $L \Delta_{\rm t}/{\pi v_s}$ 
and $L \Delta_{\rm s}/{\pi v_s}$ including finite
size corrections for $J_1=0.3$ . These are
estimated from the numerical diagonalization up to $L=10$, 
based on the size dependence of low-lying energy spectra 
predicted by the CFT.}
\end{figure}

The exponent 
$\eta$ 
can be determined from 
the triplet and singlet excitation gaps in the finite size system, 
by using the relations $\eta=L \Delta_{\rm t}/{\pi v_s}+\cdots$
and $\eta=L \Delta_{\rm s}/{\pi v_s}+\cdots$. 
Using the results from the numerical diagonalization for $L=8$ and 10, 
we estimate $c$, $L \Delta_{\rm t}/{\pi v_s}=\eta_t$ and 
$L \Delta_{\rm s}/{\pi v_s}=\eta_s$ independently, shown in 
Fig.~\ref{c-eta} for $J_1=0.3$. 
Here, we have evaluated the spin wave velocity by $v_s=(E_1-E_0)/k_1$, 
where $E_1$ is the lowest energy of the eigenstate 
with the smallest nonzero wave number $k_1=\frac{2 \pi}{L}$.
The velocity is generally nonuniversal and depends on the couplings $J_1$ and
$J_{\rm r}$.

In Fig.~\ref{c-eta}, we also depict a special average 
$(3 L \Delta_{\rm t}/{\pi v_s} +L \Delta_{\rm s}/{\pi v_s})/4$  
of the singlet and the triplet gaps 
such that the dominant finite-size logarithmic corrections cancel out 
each other.~\cite{ZS} This average and $c$ seem to be almost unity 
in the gapless phases in Fig.~\ref{c-eta} as observed in 
the $S=\frac{1}{2}$ antiferromagnetic Heisenberg chain with the 
next-nearest-neighbor interaction.~\cite{on} 
These results are completely consistent with the expected 
BKT 
transition. We therefore conclude that both gapless 
phases (II) and (IV) are governed by the WZW model,
and the transition between (II)-(III) and 
that between (III)-(IV) belong to the BKT 
universality class.

Now, we note the level crossing between $L \Delta_{\rm t}/{\pi v_s}$ and 
$L \Delta_{\rm s}/{\pi v_s}$ around $\alpha=1$ in Fig.~\ref{c-eta}. 
This level crossing implies the appearance of another
singlet ground state as a reflection of the valence-bond order 
in the thermodynamic limit. 
This is a clear numerical evidence of the extended gapful phase (III) 
predicted by several effective theories. 
On the contrary, Nishimoto and Arikawa~\cite{arikawa} have claimed that 
the gapless phase is extended everywhere
except at the point $\alpha=1$, 
by means of the DMRG analysis. 
Our observation does not agree with their claim.

\subsection{Level-spectroscopy method}
\label{level_cross}
\begin{figure}
\includegraphics[width=0.75\linewidth,angle=0]{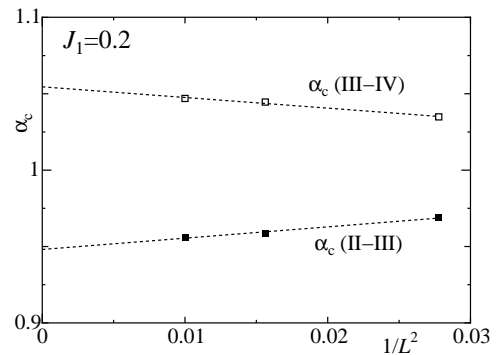}%
\caption{\label{ls1} 
Critical points $\alpha_c$ evaluated by the 
level-spectroscopy analysis for $J_1=0.2$, assuming 
the size correction is proportional to $1/L^2$. 
}
\end{figure}

The level-spectroscopy method~\cite{on,no,LS2,LS3} is 
a very powerful tool to determine the critical point of 
the BKT 
transition in one-dimensional quantum systems. 
For the SU(2)-symmetric cases including the present spin tube, its
strategy becomes easier.~\cite{on} According to this method, 
the critical point can be determined as an intersection 
between the singlet and the triplet excitation gaps, 
where their logarithmic finite-size corrections vanish.
The phase boundaries (II)-(III) and (III)-(IV) estimated by 
this method are shown in Fig.~\ref{ls1}. 
Here we have applied the results of the numerical diagonalization 
up to $L=10$ for $J_1=0.2$. 
The numerical data are well converged to those of the thermodynamic limit
by use of the $1/L^2$ extrapolation.
The $1/L^2$ extrapolation is justified by considering
the most important finite-size correction to the excitation gaps $\Delta$ next to 
the logarithmic term.\cite{on}

Several critical points estimated by the level spectroscopy are 
plotted by open circles in Fig.~\ref{phasen}. We find that 
at least for $J_1 \ll 1$, the results are in good agreement with 
the phase boundaries (II)-(III) and (III)-(IV) evaluated 
in the preceding Sec.~\ref{PRG}.

\begin{figure}
\includegraphics[width=0.75\linewidth,angle=0]{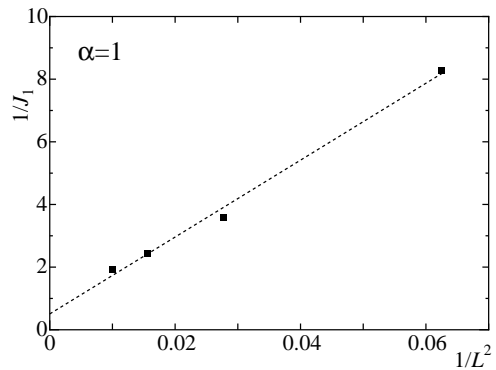}%
\caption{\label{ls2} 
Critical value of $J_1$ for the regular triangle 
spin tube ($\alpha=1$) evaluated by the level-spectroscopy analysis.}
\end{figure}

On the other hand, when we consider the transition between two phases
(III) and (I) along the line $\alpha=1$, 
the finite-size correction to the critical point is too large 
to determine the precise value of the infinite-length limit, 
as shown in Fig.~\ref{ls2}. 
The extrapolated value of the critical $J_1$ results in 
$1/J_{1c}= 0.51 \pm 0.45$. It is therefore difficult to conclude 
whether $J_{1c}$ is finite or zero for $\alpha=1$, namely, 
whether or not the gapless phase (I) still survives for small $\alpha$. 
[we have already shown the presence of the phase (I) for large
$\alpha\gg 1$ in Sec.~\ref{gapless_phase}.] 

\section{Conclusions and discussions}
\label{summary}
We have studied the ground-state phase diagram and the quantum phase 
transitions of the $S=\frac{1}{2}$ three-leg asymmetric spin tube
models, defined by Eq.~(\ref{ham}). 
In Sec.~\ref{global}, based on the Hubbard model on the tube lattice, 
we have proposed an effective theory 
to draw a global phase diagram, Fig.~\ref{phase_eff}, 
in the parameter space $(J_1, \alpha)$. 
Three gapless phases (I), (II), and (IV)
and one gapful phase (III) are found by counting the Fermi points.
This effective theory indicates that
the level-1 SU(2) WZW model describes two extended gapless 
phases (II) and (IV) which are separated by a extended gapful phase
(III) around $\alpha=1$. 
In Sec.~\ref{gapless_phase}, applying another analytical strategy 
based on the non-Abelian bosonization, 
we have proved that the predicted gapless phase (I) is 
exactly present at least for weak rung couplings $J_1 \gg J_{\rm r}$ 
with a strong asymmetry $\alpha \gg 1$. Furthermore, we have argued 
that the gapful region (III) is surely extended (although narrow)
for the case $J_{\rm r} \gg J_1$ and $J_{\rm r} \sim J_{\rm r}'$ 
in Sec.~\ref{energy_gap}.

Following these results of effective theories in Sec.~\ref{effective}, 
we have numerically analyzed the quantum phase transitions of the spin
tube~(\ref{ham}) in Sec.~\ref{numerical}. 
The phenomenological renormalization approach based on the DMRG and the
numerical diagonalization has enabled us to draw the global phase diagram. 
The numerical results are qualitatively in agreement with those of 
effective theories. 
In addition, we have raised the validity of the phase diagram by means
of the numerical finite-size scaling arguments based on the $c=1$ CFT. 
We have confirmed that the phase transitions (II)-(III) and (III)-(IV) 
belong to the BKT 
universality class quantitatively. 
Here, we make a comment on that the gapped region estimated in
Ref.~\onlinecite{arikawa}, where an unconventional power law fitting with
respect to $L$ is employed, is well inside of the present phase III. We
should, however, recall that the gap near the
BKT 
transition is exponentially small and thus the gap is difficult 
to be detected by the phenomenological renormalization 
approach, which usually  overestimates the gapless
region. This suggests that the gapped region in
Ref.~\onlinecite{arikawa} 
becomes smaller than the present phase diagram (Fig.~\ref{phasen}), 
although the DMRG data itself may be consistent with each other.

The semi-quantitative phase diagram of the spin tube~(\ref{ham}) has
been constructed in this study (see Figs.~\ref{phase_eff} and
\ref{phasen}), but some subtle issues are still remaining, e.g., 
(A) the topology of the phase boundaries along the two lines
$\alpha=0$ and $1/\alpha=0$, 
(B) how widely the gapless phase is extended, etc. 
In order to resolve these problems, 
more sophisticated approaches would be necessary. 
\\

\begin{acknowledgments}

We would like to thank M. Arikawa, J. -B. Fouet, A. L\"auchli and 
F. Mila for invaluable discussions. This work was
partly supported by Grants-in-Aid for Scientific Research (B) 
(No.$\,$17340100, and No.$\,$20340096), 
Scientific Research (C) (No.$\,$18540340), 
Priority Areas ``Invention of Anomalous Quantum Materials -New 
Physics through Innovation Materials-'' (No.$\,$19014019), 
``Physics of New Quantum Phases in Superclean Materials'' 
(No.$\,$17071011, No.$\,$18043023, and No.$\,$20029020)
and ``High Field Spin Science in 100T'' (No.$\,$20030008, and No.$\,$2003003)
from the Ministry of Education, Culture, Sports, Science, and Technology of Japan.
We further thank the Supercomputer Center, Institute
for Solid State Physics, University of Tokyo, the Cyberscience Center, Tohoku
University, and the Computer Room, Yukawa Institute for Theoretical Physics,
Kyoto University for computational facilities.

\end{acknowledgments}


\end{document}